\documentclass[a4paper,aps,prd,nofootinbib,onecolumn,notitlepage]{revtex4}
\RequirePackage[english]{babel}
\RequirePackage[latin1]{inputenc}
\RequirePackage[T1]{fontenc}
\RequirePackage{mathrsfs}
\RequirePackage{amsmath}
\RequirePackage{amssymb}
\RequirePackage{amsbsy}
\RequirePackage{bm}
\usepackage[lofdepth,lotdepth]{subfig}
\usepackage{graphicx}
\usepackage{multirow} 
\usepackage{caption}
\def\de#1/de#2{\frac{\partial {#1}}{\partial {#2}}}

\newcommand{\nn}{\nonumber}
\newcommand{\ba}{\begin{eqnarray}}
\newcommand{\ea}{\end{eqnarray}}
\newcommand{\be}{\begin{equation}}
\newcommand{\ee}{\end{equation}}

\begin{document}
\title{On the Anisotropic Interior Solutions in Ho\v rava Gravity and Einstein-\ae ther Theory}
\author{Daniele Vernieri}
\author{Sante Carloni}
\affiliation{Centro de Astrof\'isica e Gravita\c c\~ao - CENTRA,
Departamento de F\'{\i}sica, Instituto Superior T\'ecnico - IST,
Universidade de Lisboa - UL, Avenida Rovisco Pais 1, 1049-001,
Portugal}
\date{\today}
\begin{abstract}
We find a reconstruction algorithm able to generate all the static spherically symmetric interior solutions in the framework of Ho\v rava gravity and Einstein-\ae ther theory in presence of anisotropic fluids. We focus for simplicity on the case of a static \ae ther finding a large class of possible viable interior star solutions which present a very rich phenomenology. We study one illustrative example in more detail.
\end{abstract}
\maketitle


\paragraph{Introduction.} 

Ho\v rava gravity has been proposed in 2009 as an UV complete theory for quantum gravity which breaks Lorentz invariance~\cite{Horava:2009uw,Blas:2009qj} and it is expected to be renormalizable~\cite{Barvinsky:2015kil,Bellorin:2016wsl,Barvinsky:2017zlx}. An interesting feature of this theory is that it admits a covariant formulation which coincides with Einstein-\ae ther theory~\cite{Jacobson:2000xp} once the \ae ther has been chosen to be hypersurface-orthogonal at the level of the action~\cite{Jacobson:2010mx}. Ho\v rava gravity has been proven to pass all the tests at Newtonian~\cite{Blas:2009qj}, post-Newtonian~\cite{Blas:2010hb,Blas:2011zd,Bonetti:2015oda}, astrophysical (binary pulsars)~\cite{Yagi:2013qpa,Yagi:2013ava} and cosmological scales~\cite{Audren:2014hza,Frusciante:2015maa}. One big challenge in this endeavour concerns the phenomenology of the theory, such as the study of black holes~\cite{Barausse:2011pu,Blas:2011ni,Berglund:2012bu,Wang:2012nv,Barausse:2012qh,Barausse:2013nwa,Eling:2016ixk} and interior solutions for relativistic stars~\cite{Eling:2006df,Eling:2007xh}. Indeed, because of the intrinsic highly non-linear structure of its field equations, even at low-energies, it is not trivial to find exact analytical solutions without reducing to very specific cases. In this letter, we focus on exact static spherically symmetric interior solutions of the low-energy limit of the covariantized version of Ho\v rava gravity  in presence of an anisotropic fluid. These solutions can be employed to model compact objects (see {\it e.g.}~\cite{Harko:2002db} and references therein) and derive observational constraints  based on their structure. We will consider here the fully anisotropic case as it is well known that in many classes of astrophysical objects  a number of processes can generate such anisotropies ({\it e.g.} viscosity, phase transitions, etc.~\cite{Herrera:1997plx}).

In the following, we will show that in the context of the covariant formulation of Ho\v rava gravity there exists a simple reconstruction algorithm able to generate a double infinity of such anisotropic solutions. Indeed, once the structure of the interior spacetime is chosen, it will be possible to solve algebraically in a trivial way the field equations for the density, the radial and transversal pressure. Such result can be considered as a first step in the study of the full structure and the properties of this type of system in Ho\v rava gravity and the comprehension of its outstanding phenomenological implications. In fact, Ho\v rava gravity is expected to complete General Relativity (GR) at very high-energies where the latter fails in describing the spacetime structure, {\it e.g.} in the case of curvature singularities. Then, it is certainly of interest to unveal the properties of spacetimes where strong-field gravity effects are expected to dominate such as in the interior of compact objects. Moreover, since the theory breaks Lorentz invariance at all energy scales, its dynamical degrees of freedom which are a spin-2 and a spin-0 graviton, can also propagate at superluminal, even infinite, velocities. The extra scalar degree of freedom physically represents the preferred foliation of the spacetime and in this framework such a foliation is not just a mathematical artifact but it instead encodes the main features related to the causal structure predicted by the theory. So, one expects that such extra scalar gravitational mode might drastically affect the interior solutions for compact objects, eventually leaving some imprint of its existence.

\paragraph{The theory.}

The action of the low-energy limit of Ho\v rava gravity is:
\be \label{horava}
\mathcal{S}_{H}=\frac{1}{16\pi G_H}\int{dT d^3x\sqrt{-g}\left(K_{ij}K^{ij}-\lambda K^2 +\xi \mathcal{R}+\eta a_i a^i\right)}+S_m[g_{\mu\nu},\psi],
\ee 
where $G_H$ is the effective gravitational constant, $T$ is the preferred time, $g$ is the determinant of the metric $g_{\mu\nu}$, $\mathcal{R}$ is the Ricci scalar of the three-dimensional constant-$T$ hypersurfaces, $K_{ij}$ is the extrinsic curvature and $K$ is its trace, $a_i=\partial_i \mbox{ln} N$ where $N$ is the lapse function of the ADM metric and $S_m$ is the matter action for the matter fields collectively denoted by $\psi$. The constants $\left\{\lambda,\xi,\eta\right\}$ are dimensionless and in the case of GR are respectively $\left\{1,1,0\right\}$. In what follows we consider the covariantized version of the theory referred to as the {\it khronometric} theory. In this respect, let us consider the action of Einstein-\ae ther theory~\cite{Jacobson:2000xp}, that is:
\be
\mathcal{S}_{\mbox{\scriptsize\ae}}=-\frac{1}{16\pi G_{\mbox{\footnotesize \ae}}}\int{d^4x\sqrt{-g}\left(R+\mathcal{L}_{\mbox{\scriptsize\ae}}\right)}+S_m[g_{\mu\nu},\psi], \label{aether}
\ee
where $G_{\mbox{\footnotesize \ae}}$ is the ``bare'' gravitational constant, $u^a$ is a unit timelike vector field, {\it i.e.} $g_{\mu\nu}u^\mu u^\nu=1$, from now on referred to as the ``\ae ther'', and $\mathcal{L}_{\mbox{\scriptsize\ae}}$ is
\be
\mathcal{L}_{\mbox{\scriptsize\ae}}=c_1\nabla^\alpha u^\beta\nabla_\alpha u_\beta+c_2\nabla_\alpha u^\alpha\nabla_\beta u^\beta+c_3\nabla_\alpha u^\beta\nabla_\beta u^\alpha+c_4 u^\alpha u^\beta\nabla_\alpha u_\nu\nabla_\beta u^\nu\,,
\ee
where the $c_i$ are arbitrary constants. Let us take the \ae ther to be hypersurface-orthogonal at the level of the action, which locally amounts to choosing
\be
u_\alpha=\frac{\partial_\alpha T}{\sqrt{g^{\mu\nu}\partial_\mu T \partial_\nu T}}\,,
\ee
where in the covariant formulation the preferred time $T$ becomes a scalar field (the ``{\it khronon}'') which defines the preferred foliation. Then, the two actions in Eqs.~\eqref{horava} and~\eqref{aether} can be shown to be equivalent by means of the following mapping of the parameters~\cite{Jacobson:2010mx}:
\be
	\label{eqn:corresp}
\frac{G_H}{G_{\scriptsize\mbox{\ae}}}=\xi = \frac{1}{1-c_{13}}\,, \hspace{2em} \frac{\lambda}{\xi} = 1 + c_2\,, \hspace{2em} \frac{\eta}{\xi} = c_{14}\,,
\ee
where $c_{ij} = c_i+c_j$.  The independent field equations for the low-energy limit of Ho\v rava gravity can then be written in a covariant form.

\paragraph{The field equations.}

Let us now consider a spherically symmetric spacetime where the metric is written in general as:
\be
ds^2 = A(r) d t^2 - B(r) d r^2 - r^2\,\big(d\theta^2+\sin^2\theta d\phi^2\big), \label{Eq0}
\ee
with the addition of an anisotropic fluid with a stress-energy tensor given by 
\be
T_\mu\,^\nu=\mbox{diag}\big(\rho,-p_r,-p_t,-p_t\big),
\ee 
where $\rho$ is the density of the fluid, $p_r$ and $p_t$ are the radial and transversal pressure respectively. Let us take into account, for simplicity, a static \ae ther $u^\mu$~\cite{Eling:2006df,Eling:2007xh}
\be
u^\mu =\biggl(\frac{1}{\sqrt{A}},0,0,0\biggr),
\ee
which is always hypersurface-orthogonal.
With the choices of the spherically symmetric background, the anisotropic fluid and the static \ae ther written above, the field equations reduce to:
\be
\frac{\eta }{ \xi  }\left[-\frac{A''(r)}{2  A(r)^2 B(r)}+\frac{ A'(r) B'(r)}{4  A(r)^2 B(r)^2}+\frac{3   A'(r)^2}{8  A(r)^3 B(r)}-\frac{ A'(r)}{ r A(r)^2 B(r)}\right]+\frac{B'(r)}{r A(r) B(r)^2}-\frac{1}{r^2 A(r) B(r)}+\frac{1}{r^2 A(r)}=\frac{8 \pi G_{\mbox{\footnotesize \ae}} \rho (r)}{A(r)}\,, \label{Eq1}
\ee 
\be
\frac{\eta A'(r)^2}{8 \xi  A(r)^2 B(r)^2}+\frac{A'(r)}{r A(r) B(r)^2}+\frac{1}{r^2 B(r)^2}-\frac{1}{r^2 B(r)}=\frac{8 \pi G_{\mbox{\footnotesize \ae}} p_r(r)}{B(r)}\,, \label{Eq2}
\ee
\be
-\frac{\eta  A'(r)^2}{8 \xi  r^2 A(r)^2 B(r)}+\frac{A''(r)}{2 r^2 A(r) B(r)}-\frac{A'(r) B'(r)}{4 r^2 A(r) B(r)^2}+\frac{A'(r)}{2 r^3 A(r) B(r)}-\frac{A'(r)^2}{4 r^2 A(r)^2 B(r)}-\frac{B'(r)}{2 r^3 B(r)^2}=\frac{8 \pi G_{\mbox{\footnotesize \ae}} p_t(r)}{r^2}\,, \label{Eq3}
\ee
plus the conservation equation for the stress-energy tensor of matter:
\be
p_r'(r)+\left[\rho(r)+p_r(r)\right]\frac{A'(r)}{2A(r)}=\frac{2}{r} \left[p_t(r)-p_r(r)\right]. \label{conserv}
\ee
In the above equations, the effective density and pressures contributions of the \ae ther are the ones controlled by the parameter $\eta/\xi$.
GR is automatically recovered from the equations above when $\eta=0$. However, one does not need to consider all of the above 4 equations, but only 3, since they are related to each other. Indeed, it can be easily shown that, by expressing $\rho(r)$, $p_r(r)$ and $p_t(r)$ in terms of $A(r)$, $B(r)$ and their derivatives respectively from Eqs.~\eqref{Eq1},~\eqref{Eq2} and~\eqref{Eq3}, and substituting them into the conservation Eq.~\eqref{conserv}, then the latter is trivially satisfied. It should be pointed out that this result only holds in the case of a static \ae ther since, otherwise, two more equations (which in this case are identically zero) should be taken into account. It is also worth noticing that, in spite of the differences in the general field equations, in our specific case (spherical symmetry, static \ae ther) the above equations coincide exactly with the ones one would obtain in Einstein-\ae ther theory~\cite{Blas:2010hb}.

\paragraph{A reconstruction algorithm.} 

The structure of the field equations above suggests an easy procedure to generate exact solutions in this context. Indeed, once the metric coefficients $A(r)$ and $B(r)$ have been chosen, the density $\rho(r)$ and the pressures $p_r(r)$ and $p_t(r)$ can be algebraically solved by means respectively of the Eqs.~\eqref{Eq1},~\eqref{Eq2} and~\eqref{Eq3} as: 
\be
\rho (r) = \frac{1}{8 \pi G_{\mbox{\footnotesize \ae}}} \left[-\frac{\eta A''(r)}{2 \xi  A(r) B(r)}+\frac{\eta  A'(r) B'(r)}{4 \xi  A(r) B(r)^2}+\frac{3 \eta  A'(r)^2}{8 \xi  A(r)^2 B(r)}-\frac{\eta  A'(r)}{\xi  r A(r) B(r)}+\frac{B'(r)}{r B(r)^2}-\frac{1}{r^2 B(r)}+\frac{1}{r^2}\right]\,, \label{Eq4}
\ee
\be
p_r(r) = \frac{1}{8 \pi G_{\mbox{\footnotesize \ae}}} \left[\frac{\eta A'(r)^2}{8 \xi  A(r)^2 B(r)}+\frac{A'(r)}{r A(r) B(r)}+\frac{1}{r^2 B(r)}-\frac{1}{r^2}\right]\,, \label{Eq5}
\ee
\be
p_t(r) = \frac{1}{8 \pi G_{\mbox{\footnotesize \ae}}} \left[\frac{A''(r)}{2 A(r) B(r)}-\frac{A'(r) B'(r)}{4 A(r) B(r)^2}+\frac{A'(r)}{2 r A(r) B(r)}-\frac{\eta  A'(r)^2}{8 \xi A(r)^2 B(r)}-\frac{A'(r)^2}{4 A(r)^2 B(r)}-\frac{B'(r)}{2 r B(r)^2}\right]\,. \label{Eq6}
\ee
We are therefore left with a double infinity of exact analytical solutions which can be studied in detail. 
Such solutions, in order to represent the interior of realistic astrophysical objects, should satisfy very specific physical conditions (see {\it e.g.}~\cite{Harko:2002db}). Indeed, the curvature invariants $R$, $R_{\mu\nu}R^{\mu\nu}$ and $R_{\mu\nu\gamma\sigma}R^{\mu\nu\gamma\sigma}$ must be smooth and finite, the ther\-mo\-di\-na\-mi\-cal quantities $\rho$, $p_r$ and $p_t$ must be smooth, finite and positive decreasing functions anywhere in the interior region, and the pressures should assume the same value in the centre. Also, $p_r$ has to be zero at the surface of the star in order to guarantee the stability of the spherical object and the implementation of the junction conditions~\cite{Israel:1966rt} to the exterior vacuum metric. Finally, one should also require that the fluid only has subluminal propagation speeds in the radial and transversal direction, {\it i.e.} the squared radial and transversal speed of sound, respectively $c_r^2=\frac{d p_r}{d\rho}$ and $c_t^2=\frac{d p_t}{d\rho}$, can only take values between $0$ and $1$. 

Notice that, all of the above requirements are just necessary but not sufficient in order to guarantee the stability of a physical object, for which only a perturbative analysis might give a definitive answer. Moreover, we are not taking into account here any specific equation of state for the fluid. The reason for this is twofold: on the one hand, even in GR the relativistic equation of state describing the interior of a star is unknown at the present time and there is a large variety of theoretical approaches that are used to find its effective description (see~\cite{Ozel:2016oaf;Hernandez:1998tz} and references therein). Secondly, in a theory like the one under scrutiny, the physical properties of the fluid might be drastically different with respect to the ones usually considered in the standard scenario. For these reasons, taking into account a specific model for the fluid would just be at the level of wishful thinking at this stage. Then, for the purpose of the current paper, we leave the equation of state unspecified.

We will now employ this reconstruction algorithm to generate a specific solution which, as we will show, can be relevant as a model for the interior of a relativistic star. We will also briefly analyse some of its physical properties. In passing, it is somewhat curious that solving for an isotropic spacetime ($p_r =p_t$) leads to an additional differential constraint. This fact makes the search for exact isotropic solutions considerably more difficult than their anisotropic counterparts, in spite of the fact that the latter are more physically relevant.

\paragraph{An illustrative example.}

Let us consider the following choice of the metric coefficients:
\be
A(r)=D_1+D_2 r^2\,, \,\,\,\,\,\, B(r)=\frac{D_3+D_4 r^2}{D_3+D_5 r^2+D_6 r^4}\,. \label{Eq4}
\ee
Notice that the metric coefficients above are qualitatively the same as the ones characterizing the Tolmann IV solution for an isotropic fluid~\cite{Tolman:1939jz}, from which they differ only for the additional constants appearing in $B(r)$. The choice of $A(r)$ is motivated by the fact that it reproduces the Newtonian potential for a fluid sphere of constant density. Instead $B(r)$ has been chosen for convenience in the calculations. The choice of the constants in its expression guarantees the avoidance of a divergence in the centre for the curvature invariants $R$, $R_{\mu\nu}R^{\mu\nu}$ and $R_{\mu\nu\gamma\sigma}R^{\mu\nu\gamma\sigma}$ and both for $\rho(r)$ and $p_r(r)$.
Applying the reconstruction algorithm, the thermodynamical quantities $\rho(r)$, $p_r(r)$ and $p_t(r)$ corresponding to the metric~\eqref{Eq4} are:
\ba
\rho(r) &=& -\frac{1}{16 \pi G_{\mbox{\footnotesize \ae}} \xi \left(D_1+D_2 r^2\right)^2 \left(D_3+D_4 r^2\right)^2}\left\{-2 D_1^2 \xi  \left[D_3 \left(3 D_4-3 D_5-5 D_6 r^2\right)+D_4 r^2 \left(D_4-D_5-3 D_6 r^2\right)\right]\right. \nn \\
&&+2 D_1 D_2 \left[3 D_3^2 \eta +D_3 r^2 \left(2 D_4 (\eta -3 \xi )+D_5 (4 \eta +6 \xi )+5 D_6 r^2 (\eta +2 \xi )\right)+D_4 r^4 \left(-2 D_4 \xi +D_5 (3 \eta +2 \xi ) \right.\right. \label{Eq7} \nn \\
&&\left.\left.+2 D_6 r^2 (2 \eta +3 \xi )\right)\right]+D_2^2 r^2 \left[3 D_3^2 \eta +D_3 r^2 \left(D_4 (\eta -6 \xi )+D_5 (5 \eta +6 \xi )+D_6 r^2 (7 \eta +10 \xi )\right) \right. \nn \\
&&\left.\left. +D_4 r^4 \left(-2 D_4 \xi +D_5 (3 \eta +2 \xi )+D_6 r^2 (5 \eta +6 \xi )\right)\right]\right\}, \label{Eq8} \nn \\
\ea
\ba
p_r(r)&=&\frac{1}{16 \pi G_{\mbox{\footnotesize \ae}} \xi \left(D_1+D_2 r^2\right)^2 \left(D_3+D_4 r^2\right)} \left\{2 D_1^2 \xi  \left(-D_4+D_5+D_6 r^2\right)+4 D_1 D_2 \xi \left[D_3+r^2 \left(2 \left(D_5+D_6 r^2\right)-D_4\right)\right]\right. \nn \\
&&\left.+D_2^2 r^2 \left[D_3 (\eta +4 \xi )+r^2 \left(-2 D_4 \xi +D_5 (\eta +6 \xi )+D_6 r^2 (\eta +6 \xi )\right)\right]\right\}, \label{Eq9} \nn \\
\ea
\ba
p_t(r)&=&\frac{1}{16 \pi G_{\mbox{\footnotesize \ae}} \xi \left(D_1+D_2 r^2\right)^2 \left(D_3+D_4 r^2\right)^2} \left\{2 D_1^2 \xi  \left[D_3 \left(-D_4+D_5+2 D_6 r^2\right)+D_4 D_6 r^4\right]+2 D_1 D_2 \xi  \left[2 D_3^2 \right.\right. \nn \\
&&\left. +D_3 r^2 \left(-D_4+5 D_5+8 D_6 r^2\right)+D_4 r^4 \left(2 D_5+5 D_6 r^2\right)\right]-D_2^2 r^2 \left[D_3^2 (\eta -2 \xi)+D_3 r^2 \left(D_4 (\eta +2 \xi) \right.\right. \nn \\
&&\left.\left.\left. +D_5 (\eta -6 \xi )+D_6 r^2 (\eta -10 \xi )\right)+D_4 r^4 \left(D_5 (\eta -2 \xi )+D_6 r^2 (\eta -6 \xi )\right)\right]\right\}, \nn \\
\ea
where $D_1$, $D_2$, $D_3$, $D_4$, $D_5$ and $D_6$ are arbitrary constants. \\
A solution like the one above will represent a stellar object only if some specific conditions are satisfied~\cite{Harko:2002db}. We will briefly check if there exists at least a combination of the parameters in Eq.~\eqref{Eq4} for which these conditions are satisfied. In the centre, $r=0$, the density $\rho$ is finite being
\be
\rho(0)=\frac{3 \left[D_1 \xi  (D_4-D_5)-D_2 D_3 \eta\right]}{8 \pi G_{\mbox{\footnotesize \ae}} \xi D_1 D_3}\,,
\ee
and both $p_r$ and $p_t$ are finite as well and assume the same value which is
\be
p_r(0)=p_t(0)=\frac{D_1 D_5-D_1 D_4+2 D_2 D_3}{8 \pi G_{\mbox{\footnotesize \ae}}  D_1 D_3}\,.
\ee
Moreover, the curvature invariants $R$, $R_{\mu\nu}R^{\mu\nu}$ and $R_{\mu\nu\gamma\sigma}R^{\mu\nu\gamma\sigma}$ are smooth and finite all over the interior region and even in the centre.
Asking the continuity of $A(r)$, $B(r)$ and $A'(r)$ at $r=\bar{R}$, that corresponds to the implementation of the junction conditions~\cite{Israel:1966rt} to the exterior vacuum metric, from Eq.~\eqref{Eq2} we read that also $p_r(r)$ has to be continuous at the surface $r=\bar{R}$, that is $p_r(\bar{R})=0$. The latter, by using Eq.~\eqref{Eq9}, gives the following relation among the constants and the radius $\bar{R}$ of the star:
\ba
D_3 = \frac{2 D_1^2 \xi  \left(D_4-D_5-D_6 \bar{R}^2\right)+4 D_1 D_2 \xi \bar{R}^2 \left[D_4-2 \left(D_5+D_6 \bar{R}^2\right)\right]-D_2^2 \bar{R}^4 \left[-2 D_4 \xi +D_5 (\eta +6 \xi )+D_6 \bar{R}^2 (\eta +6 \xi )\right]}{D_2 \left[4 D_1 \xi +D_2 \bar{R}^2 (\eta +4 \xi )\right]}\,. \nn \\
\ea
If one intends to guarantee a smoother behaviour of the scalar curvature at the star surface, one could require also the continuity of $B'(r)$ across the boundary. While in GR this would force $\rho(r)$ to be continuous, that is $\rho(\bar{R})=0$, in the case of Ho\v rava gravity Eq.~\eqref{Eq1} implies that $A''(\bar{R})$ determines the jump of $\rho(r)$ on the boundary, so that we can have $\rho(\bar{R})\neq0$. 
\begin{figure}[]
\includegraphics[width=0.5\textwidth]{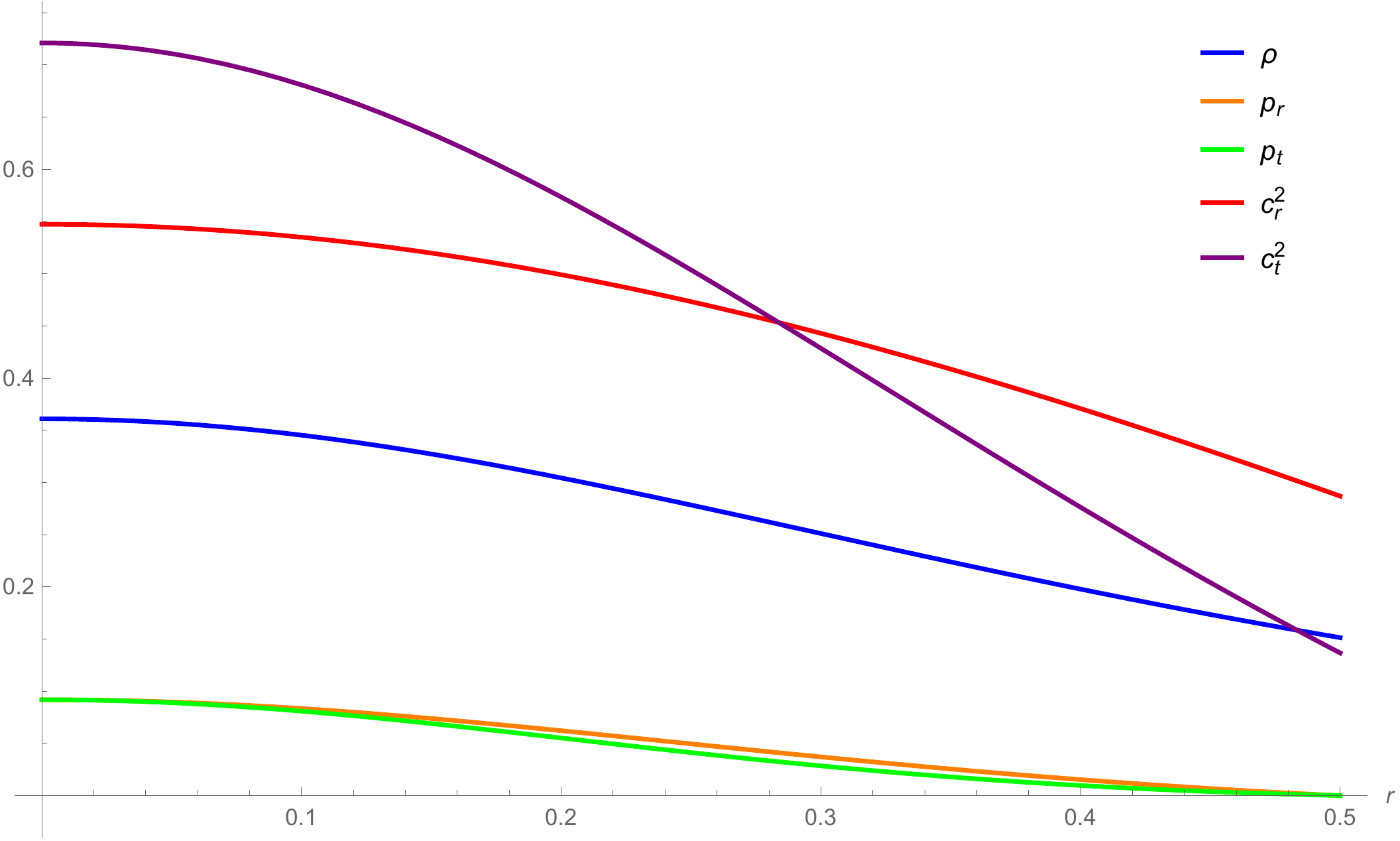}
\caption{The figure shows the behaviour of the density $\rho$ (blue line), the radial pressure $p_r$ (orange line), the tangential pressure $p_t$ (green line), the squared radial speed of sound $c_r^2$ (red line) and the squared tangential speed of sound $c_t^2$ (purple line) as functions of the radius for a wide choice of the constants in units $G_N=1$. In this specific case the constants take the values: $\xi=1.00001$, $D_1=-30$, $D_2=-80$, $D_4=-50$, $D_5=10$, $D_6=-10$, $\bar{R}=0.5$.}
\label{Figure1}
\end{figure}
The lack of a general explicit vacuum solution in Ho\v rava gravity, even with a static \ae ther~\cite{Eling:2006df,Eling:2007xh}, does not make it possible at this stage to impose explicitly the continuity conditions at the surface written above, and then, to eventually find a relation between the radius and the mass of the star. Also, in the case of a static \ae ther under study, the Schwarzschild metric is a vacuum solution only if $\eta=0$ identically, which would collapse Eqs.~\eqref{Eq1}-\eqref{Eq3} to the ones of GR. 
In Fig.~\ref{Figure1} we show the behaviour of $\rho$, $p_r$, $p_t$, $c_r^2$ and $c_t^2$ as a function of the radius $r$ for some choice of the constants $D_i$, with $G_{\mbox{\footnotesize \ae}}=G_N \left(1-\eta/2\xi \right)$, where $G_N$ is the Newton's constant, so that the Newtonian limit is recovered~\cite{Carroll:2004ai}, and with $\eta = 2\left(\xi-1\right)$ which is required to evade the post-Newtonian constraints~\cite{Blas:2010hb,Blas:2011zd,Bonetti:2015oda}. As it is clear from Fig.~\ref{Figure1}, all of the aforementioned physical quantities are finite and positive decreasing functions from the center up to the surface $r=R$ where $p_r=0$, then they are well behaving in describing the interior of compact objects. Moreover, both $c_r^2$ and $c_t^2$ take values between $0$ and $1$, which means that the fluid only has subluminal propagation speeds in the two different directions. Finally, notice from Fig.~\ref{Figure1} that in the case under study the fluid shows a quasi-isotropic behaviour because of the very small difference between $p_r(r)$ and $p_t(r)$, which makes the anisotropy very near to zero.

\paragraph{Conclusions.}

The purpose of this letter has been to propose a reconstruction algorithm to explore exact analytical static spherically symmetric interior solutions in the framework of the covariantized  version of Ho\v rava gravity and in presence of anisotropic fluids and a static \ae ther. Indeed, since in Eqs.~\eqref{Eq2}-\eqref{Eq3} the effective pressures generated by the \ae ther differ only by a sign, a static \ae ther gives naturally rise, even in vacuum, to anisotropic contributions to the field equations. Then, it seems natural to consider in this framework interior solutions where the pressure behaves anisotropically, {\it i.e.} it acquires two different components, one radial and the other transversal. We have discussed that, due to the structure of the field equations, once the metric is assigned, the density and the two pressures are determined algebraically. Hence, a double infinity of solutions does exist, in the sense that assigning arbitrarily the two coefficients of the metric the field equations get automatically satisfied. This result is relevant since, as already mentioned, the lack of exact analytical solutions has been one of the major problems in order to understand the phenomenological implications of the theory with explicit solutions at hand. Notice that, because of the complete equivalence between the field equations of Ho\v rava gravity and of Einstein-\ae ther theory in spherical symmetry with a static \ae ther, the reconstruction algorithm that we have proposed applies in the same way in both cases. To conclude, it should be stressed out that the presence of a generating algorithm doesn't mean that all the solutions found with this technique have a physical meaning. Indeed, as we have already discussed, this should be checked case by case in order to prove that the solution at hand satisfies all the physical requirements. Notice that we are not enforcing a priori an equation of state for the interior fluid, since its form, even in GR, is still unknown. And the determination of the latter is anyway far from the scope of this paper. Moreover, one might also require that the interior solution is stable under perturbations, which is out of the scope of the present study too, and that it can be joined with a suitable exterior solution. Regarding the latter point, not many such solutions are known at present in Ho\v rava gravity and Einstein-\ae ther theory. The most general one for the case of static \ae ther was found in Refs.~\cite{Eling:2006df,Eling:2007xh}. Unfortunately such solution is in implicit form and can only be analytically inverted  for specific values of $\eta$ and $\xi$ which are incompatible with the measured value of the post-Newtonian parameters. This implies that the only way in which the junction could be performed is via numerical analysis. A numerical approach defies the very purpose of an analytical treatment of the problem of compact objects  that characterises the present work and for this reason we will not pursue it here. In a following work we will provide a more detailed study of the solutions generated by the reconstruction algorithm and the consequences of their junction to exterior solutions.

\vspace{0.5cm}
{\bf Acknowledgments:} DV and SC were supported by the Funda\c{c}\~{a}o para a Ci\^{e}ncia e Tecnologia through project IF/00250/2013 and acknowledge financial support provided under the European Union's H2020 ERC Consolidator Grant ``Matter and strong-field gravity: New frontiers in Einstein's theory'' grant agreement No. MaGRaTh646597.



\begin{thebibliography}{99}

\bibitem{Horava:2009uw} 
  P.~Horava,
  Phys.\ Rev.\ D {\bf 79}, 084008 (2009)
  doi:10.1103/PhysRevD.79.084008
  [arXiv:0901.3775 [hep-th]].

\bibitem{Blas:2009qj} 
  D.~Blas, O.~Pujolas and S.~Sibiryakov,
  Phys.\ Rev.\ Lett.\  {\bf 104}, 181302 (2010)
  doi:10.1103/PhysRevLett.104.181302
  [arXiv:0909.3525 [hep-th]].

\bibitem{Barvinsky:2015kil} 
  A.~O.~Barvinsky, D.~Blas, M.~Herrero-Valea, S.~M.~Sibiryakov and C.~F.~Steinwachs,
  Phys.\ Rev.\ D {\bf 93}, no. 6, 064022 (2016)
  doi:10.1103/PhysRevD.93.064022
  [arXiv:1512.02250 [hep-th]].
	
\bibitem{Bellorin:2016wsl} 
  J.~Bellor\'in and A.~Restuccia,
  Phys.\ Rev.\ D {\bf 94}, no. 6, 064041 (2016)
  doi:10.1103/PhysRevD.94.064041
  [arXiv:1606.02606 [hep-th]].
	
\bibitem{Barvinsky:2017zlx} 
  A.~O.~Barvinsky, D.~Blas, M.~Herrero-Valea, S.~M.~Sibiryakov and C.~F.~Steinwachs,
  arXiv:1705.03480 [hep-th].

\bibitem{Jacobson:2000xp} 
  T.~Jacobson and D.~Mattingly,
  Phys.\ Rev.\ D {\bf 64}, 024028 (2001)
  doi:10.1103/PhysRevD.64.024028
  [gr-qc/0007031].

\bibitem{Jacobson:2010mx} 
  T.~Jacobson,
  Phys.\ Rev.\ D {\bf 81}, 101502 (2010)
  Erratum: [Phys.\ Rev.\ D {\bf 82}, 129901 (2010)]
  doi:10.1103/PhysRevD.82.129901, 10.1103/PhysRevD.81.101502
  [arXiv:1001.4823 [hep-th]].

\bibitem{Blas:2010hb} 
  D.~Blas, O.~Pujolas and S.~Sibiryakov,
  JHEP {\bf 1104}, 018 (2011)
  doi:10.1007/JHEP04(2011)018
  [arXiv:1007.3503 [hep-th]].
	
\bibitem{Blas:2011zd} 
  D.~Blas and H.~Sanctuary,
  Phys.\ Rev.\ D {\bf 84}, 064004 (2011)
  doi:10.1103/PhysRevD.84.064004
  [arXiv:1105.5149 [gr-qc]].

\bibitem{Bonetti:2015oda} 
  M.~Bonetti and E.~Barausse,
  Phys.\ Rev.\ D {\bf 91}, 084053 (2015)
  Erratum: [Phys.\ Rev.\ D {\bf 93}, 029901 (2016)]
  doi:10.1103/PhysRevD.91.084053, 10.1103/PhysRevD.93.029901
  [arXiv:1502.05554 [gr-qc]].

\bibitem{Yagi:2013qpa} 
  K.~Yagi, D.~Blas, N.~Yunes and E.~Barausse,
  Phys.\ Rev.\ Lett.\  {\bf 112}, no. 16, 161101 (2014)
  doi:10.1103/PhysRevLett.112.161101
  [arXiv:1307.6219 [gr-qc]].
	
\bibitem{Yagi:2013ava} 
  K.~Yagi, D.~Blas, E.~Barausse and N.~Yunes,
  Phys.\ Rev.\ D {\bf 89}, no. 8, 084067 (2014)
  Erratum: [Phys.\ Rev.\ D {\bf 90}, no. 6, 069902 (2014)]
  Erratum: [Phys.\ Rev.\ D {\bf 90}, no. 6, 069901 (2014)]
  doi:10.1103/PhysRevD.90.069902, 10.1103/PhysRevD.90.069901, 10.1103/PhysRevD.89.084067
  [arXiv:1311.7144 [gr-qc]].

\bibitem{Audren:2014hza} 
  B.~Audren, D.~Blas, M.~M.~Ivanov, J.~Lesgourgues and S.~Sibiryakov,
  JCAP {\bf 1503}, no. 03, 016 (2015)
  doi:10.1088/1475-7516/2015/03/016
  [arXiv:1410.6514 [astro-ph.CO]].

\bibitem{Frusciante:2015maa} 
  N.~Frusciante, M.~Raveri, D.~Vernieri, B.~Hu and A.~Silvestri,
  Phys.\ Dark Univ.\  {\bf 13}, 7 (2016)
  doi:10.1016/j.dark.2016.03.002
  [arXiv:1508.01787 [astro-ph.CO]].

\bibitem{Barausse:2011pu} 
  E.~Barausse, T.~Jacobson and T.~P.~Sotiriou,
  Phys.\ Rev.\ D {\bf 83}, 124043 (2011)
  doi:10.1103/PhysRevD.83.124043
  [arXiv:1104.2889 [gr-qc]].

\bibitem{Blas:2011ni} 
  D.~Blas and S.~Sibiryakov,
  Phys.\ Rev.\ D {\bf 84}, 124043 (2011)
  doi:10.1103/PhysRevD.84.124043
  [arXiv:1110.2195 [hep-th]].

\bibitem{Berglund:2012bu} 
  P.~Berglund, J.~Bhattacharyya and D.~Mattingly,
  Phys.\ Rev.\ D {\bf 85}, 124019 (2012)
  doi:10.1103/PhysRevD.85.124019
  [arXiv:1202.4497 [hep-th]].

\bibitem{Wang:2012nv} 
  A.~Wang,
  Phys.\ Rev.\ Lett.\  {\bf 110}, no. 9, 091101 (2013)
  doi:10.1103/PhysRevLett.110.091101
  [arXiv:1212.1876 [hep-th]].

\bibitem{Barausse:2012qh} 
  E.~Barausse and T.~P.~Sotiriou,
  Phys.\ Rev.\ D {\bf 87}, 087504 (2013)
  doi:10.1103/PhysRevD.87.087504
  [arXiv:1212.1334 [gr-qc]].
				
\bibitem{Barausse:2013nwa} 
  E.~Barausse and T.~P.~Sotiriou,
  Class.\ Quant.\ Grav.\  {\bf 30}, 244010 (2013)
  doi:10.1088/0264-9381/30/24/244010
  [arXiv:1307.3359 [gr-qc]].

\bibitem{Eling:2016ixk} 
  C.~Eling,
  Phys.\ Rev.\ D {\bf 94}, no. 12, 126017 (2016)
  doi:10.1103/PhysRevD.94.126017
  [arXiv:1610.05967 [hep-th]].

\bibitem{Eling:2006df} 
  C.~Eling and T.~Jacobson,
  Class.\ Quant.\ Grav.\  {\bf 23}, 5625 (2006)
  Erratum: [Class.\ Quant.\ Grav.\  {\bf 27}, 049801 (2010)]
  doi:10.1088/0264-9381/23/18/008, 10.1088/0264-9381/27/4/049801
  [gr-qc/0603058].
	
\bibitem{Eling:2007xh} 
  C.~Eling, T.~Jacobson and M.~Coleman Miller,
  Phys.\ Rev.\ D {\bf 76}, 042003 (2007)
  Erratum: [Phys.\ Rev.\ D {\bf 80}, 129906 (2009)]
  doi:10.1103/PhysRevD.76.042003, 10.1103/PhysRevD.80.129906
  [arXiv:0705.1565 [gr-qc]].

\bibitem{Harko:2002db} 
  T.~Harko and M.~K.~Mak,
  Annalen Phys.\  {\bf 11}, 3 (2002)
  doi:10.1002/1521-3889(200201)11:1<3::AID-ANDP3>3.0.CO;2-L
  [gr-qc/0302104].

\bibitem{Herrera:1997plx} 
  L.~Herrera and N.~O.~Santos,
  Phys.\ Rept.\  {\bf 286}, 53 (1997).
  doi:10.1016/S0370-1573(96)00042-7

\bibitem{Israel:1966rt} 
  W.~Israel,
  Nuovo Cim.\ B {\bf 44S10}, 1 (1966)
  [Nuovo Cim.\ B {\bf 44}, 1 (1966)]
  Erratum: [Nuovo Cim.\ B {\bf 48}, 463 (1967)].
  doi:10.1007/BF02710419, 10.1007/BF02712210

\bibitem{Ozel:2016oaf;Hernandez:1998tz} 
  F.~Ozel and P.~Freire,
  Ann.\ Rev.\ Astron.\ Astrophys.\  {\bf 54}, 401 (2016)
  doi:10.1146/annurev-astro-081915-023322
  [arXiv:1603.02698 [astro-ph.HE]];
  H.~Hernandez, L.~A.~Nunez and U.~Percoco,
  Class.\ Quant.\ Grav.\  {\bf 16}, 871 (1999)
  doi:10.1088/0264-9381/16/3/017
  [gr-qc/9806029].

\bibitem{Tolman:1939jz} 
  R.~C.~Tolman,
  Phys.\ Rev.\  {\bf 55}, 364 (1939).
  doi:10.1103/PhysRev.55.364
  
\bibitem{Carroll:2004ai} 
  S.~M.~Carroll and E.~A.~Lim,
  Phys.\ Rev.\ D {\bf 70}, 123525 (2004)
  doi:10.1103/PhysRevD.70.123525
  [hep-th/0407149].
																		
\end{thebibliography}
\end{document}